\documentclass[aps,prd,showpacs,twocolumn,superscriptaddress,nofootinbib]{revtex4}
\usepackage{slashed}
\usepackage{amsfonts,graphicx,bm}
\usepackage{amsmath}
\usepackage{amssymb}

\def\met{\displaystyle{\not}E_T}
\let\jnfont=\rm
\def\NPB#1,{{\jnfont Nucl.\ Phys.\ B }{\bf #1},}
\def\PLB#1,{{\jnfont Phys.\ Lett.\ B }{\bf #1},}
\def\EPJC#1,{{\jnfont Eur.\ Phys.\ Jour.\ C }{\bf #1},}
\def\PRD#1,{{\jnfont Phys.\ Rev.\ D }{\bf #1},}
\def\PRL#1,{{\jnfont Phys.\ Rev.\ Lett.\ }{\bf #1},}
\def\MPLA#1,{{\jnfont Mod.\ Phys.\ Lett.\ A }{\bf #1},}
\def\JPG#1,{{\jnfont J.\ Phys.\ G }{\bf #1},}
\def\CTP#1,{{\jnfont Commun.\ Theor.\ Phys.\ }{\bf #1},}

\begin{document}

\title{Top quark spin and $Htb$ interaction in charged Higgs and top quark associated production  at LHC}

\author{Xue Gong}  \email{gongxue@mail.sdu.edu.cn}
\affiliation{School of Physics, Shandong University, Jinan, Shandong 250100,  China}

\author{Zong-Guo Si}  \email{zgsi@sdu.edu.cn}
\affiliation{School of Physics, Shandong University, Jinan,
Shandong 250100,  China}
\affiliation{Center for High Energy
Physics, Peking University, Beijing 100871, China}

\author{Shuo Yang}  \email{yangshuo@dlu.edu.cn}
\affiliation{Physics Department, Dalian University, Dalian, 116622, China}

\author{Ya-juan Zheng} \email{yjzheng218@gmail.com}

\affiliation{CASTS, CTS and Department of Physics, National Taiwan University, Taipei, China}

\affiliation{School of Physics, Shandong University, Jinan,
Shandong 250100,  China}

\begin{abstract}

We study the charged Higgs production at LHC via its associated production with
top quark.  The kinematic cuts are optimized to suppress the background processes
so that the reconstruction of the charged Higgs and top quark is possible.
The angular distributions with respect to top quark spin
are explored to study the $Htb$ interaction at LHC.

\end{abstract}
\maketitle

\section{Introduction}

The Standard Model (SM) of particle physics, with great success, is based on two cornerstones: gauge symmetry and 
electroweak spontaneous symmetry breaking mechanism(EWSB). The gauge group $SU(3)_{C}\times SU(2)_{L}\times U(1)_{Y}$ 
of the SM has been confirmed by the discovery of $W/Z$ bosons and lots of precision measurements.
As the other cornerstone, the mechanism of EWSB is implemented by introducing only 
one complex Higgs doublet $\Phi$ in 
the SM and then triggering EWSB after the neutral component 
of $\Phi$ developing a vacuum 
expectation value (vev). In the meanwhile, the masses of weak gauge bosons and fermions are generated. 
In the SM, there is 
only one physical neutral Higgs boson $H$ after EWSB. The discovery of the Higgs boson will 
help to unveil the mysteries 
of EWSB and mass generation of SM particles. Recently, one Higgs-like particle around 126 
GeV has been discovered at LHC 
by ATLAS and CMS collaborations\cite{Higgs_ATLAS,Higgs_CMS}. It is important to further confirm the identity of this 
particle. In the SM, only one complex scalar doublet is introduced based on the "minimal principle". It is natural to consider 
more complex scalars, for example, the two Higgs doublet structure. Especially, there are many motivations to study 
the two Higgs doublet model (2HDM). Such as, in the supersymmetric models, a single Higgs doublet is unable to give mass 
simultaneously to the charge 2/3 and charge -1/3 quarks and the anomaly cancellation also require an additional Higgs doublet. 
Another motivation for 2HDM is that it could generate a baryon asymmetry of the universe of sufficient size. 
Interestingly, ATLAS and CMS announced that there is an enhancement in the di-photon channel of Higgs 
decay $h\to \gamma\gamma$ \cite{Higgs_ATLAS, Higgs_CMS}. 
This enhancement can also be explained by charged Higgs from 2HDM\cite{comprehensive}.

There are many scenarios in 2HDM structure \cite{higgs hunter's guide, 2hdm_review}. Without imposing any discrete symmetry, 
the 2HDM suffers serious flavour changing
neutral currents (FCNC) at tree-level. For the suppression of leading order FCNC as well as CP violation in the Higgs sector, 
we consider CP-conserving 2HDMs with extra discrete symmetry. Popular Type I and II 2HDMs belong to this kind. As one of the 
minimal extensions of the SM, 2HDMs \cite{higgs hunter's guide, 2hdm_review},
has five physical Higgs scalars after the spontaneous symmetry breaking,
i.e., two neutral CP-even bosons $h_{0}$ and $H_{0}$,
one neutral CP-odd boson $A$, and two charged bosons $H^{\pm}$.
In diverse models, different scalar multiplets and singlets could generate neutral scalars and there exists mixing between
neutral scalars which makes it difficult to unentangle the Higgs properties and confirm the existence of extended Higgs 
sector. However, the discovery of the charged Higgs boson could provide an unambiguous signature of the extended Higgs
sector and help to further distinguish from different models.

Motivated by above reasons, the charged Higgs $H^{\pm}$ has been searched for  at colliders in recent years.
One model-independent direct limit is from the LEP experiments gives $M>78.6$ GeV at 95\% C.L.,
where $M$ represents the mass of charged Higgs 
by exclusive decay channels of $H^{+}\to c\bar{s}$ and $H^{+}\to\tau^{+}\nu$\cite{LEP Higgs}.
At hadron colliders, the search approaches for the charged Higgs are different in low mass range 
$M<m_t$ and in large mass range $M>m_t$. In the low mass range $M<m_t-m_b$, 
the search for the signal mainly focus on the top quark decay $t\to H^+b$ followed by decay mode $H^+\to \bar{\tau}\nu$. 
On the other hand, for large mass range $M>m_t+m_b$, the signal is from the dominant production process, 
the $gb$ fusion process $gb\to tH^-$, followed by dominant decay modes $H^- \to t\bar{b}$ and $H^-\to \tau \bar{\nu}$. The Tevatron
has put a constraint on 2HDM in the small $\tan \beta$ and large $\tan \beta$ regions for a charged Higgs boson with mass up 
to $\sim 160$ GeV \cite{Tevatron}.  
In addition, the indirect constraints can be
extracted from B-meson decays since the charged Higgs contributes to the FCNC at one loop level. 
In Type II 2HDM, a limit on the charged Higgs mass $M>316~{\rm GeV}$ at $95\%$ C.L. 
is obtained dominantly from $b\to s\gamma$ branching ratio measurement irrespective of the value of 
$\tan\beta$ \cite{typeII_constraints}.
However, in Type III or general 2HDM the phases of the Yukawa couplings are free parameters so that
$M$ can be as low as 100 GeV\cite{btosgamma1}. For more detailed discussions on phenomenological constraints
on charged Higgs, we refer to Ref. \cite{PDG2012}.

Along with the experimental search for the charged Higgs boson, extensive phenomenological studies on charged Higgs
boson production have been carried out 
\cite{collider higgs,collider higgs2,dominate process,tripleb,fourb,pt,dominate process3,Kidonakis:2004ib}.
Especially, the $gb$ fusion process $gb\rightarrow tH^{-}$
for $M>m_{t}+m_{b}$ \cite{dominate process,tripleb,fourb,pt,dominate process3,Kidonakis:2004ib,H+Jetsub}
has drawn more attentions due to the large couplings of $Htb$ interaction. 
The next-to-leading QCD corrections to this process has also been performed in order to make the 
theoretical predictions more reliable\cite{NLO}.

In this work, we revisit this process at the LHC and take a method similar to 
Ref. \cite{Gopalakrishna:2010xm} to 
distinguish the signal from backgrounds. As demonstrated in Ref. \cite{Gopalakrishna:2010xm},
the angular distribution related to top quark spin is efficient to
disentangle the chiral coupling of the $W'$ boson to SM fermions. 
The left-right asymmetry induced by top
quark spin for $pp\to tH^-$ process has been analyzed in  Ref.\cite{Baglio:2011ap}.
Here we further investigate this kind of effects after including the decay information 
of charged Higgs and top quark, and we employ the angular distributions of the top
 quark and the lepton resulting from top and 
charged Higgs decay to disentangle $Htb$ couplings at LHC. 

This paper is organized as follows. In Section \ref{secii}., the corresponding
theoretical framework is briefly introduced. Section \ref{seciii}.
is devoted to the numerical analysis of top quark and charged Higgs
associated production. Specifically, the correlated angular distributions are
investigated to identify the interaction of top-bottom quark and charged
Higgs. Finally, a short summary is given.

\section{Theoretical Framework}

\label{secii}

\subsection{Lagrangian related to the interaction of Higgs and quarks }

We start with a brief introduction to the two-Higgs-Doublet Model (2HDM) which is  one of the minimal 
extensions of SM. Different from SM, 2HDM involves two complex ${SU(2)_{L}}$ doublet scalar fields.
\begin{equation}
\Phi_{i}=\left(\begin{array}{c}
H_{i}^{+}\\
(H_{i}^{0}+iA_{i}^{0})/\sqrt{2}\end{array}\right),\end{equation}
 where $i=1, 2$. Imposing CP invariance and ${\rm U(1)_{EM}}$ gauge
symmetry, the minimization of potential gives
\begin{eqnarray}
\langle \Phi_{i}\rangle=\displaystyle{\frac{1}{\sqrt{2}}}\left(\begin{array}{c}
0\\
v_{i}\end{array}\right),\end{eqnarray}
with $v_{i}~(i=1, 2)$ is non-zero vev. One important parameter
in 2HDM ${\displaystyle {\tan\beta\equiv{v_{2}}/{v_{1}}}}$ is defined accordingly, which determines
the interactions of the various Higgs fields with the vector bosons
and fermions, it has substantial meaning in discussing
phenomenology. The most severe constraints on $\tan\beta$ and $M$ come from flavour physics including $B$ and $D$ mesons, $\Delta M_{B_d}$, $b\to s\gamma$ and $Z\to b\bar{b}$\cite{typeII_constraints, 2HDM_constraints}. Large $\tan\beta$ is favored by $B$ meson rare decays\cite{Hewett:1992is,Barger:1992dy,Bertolini:1990if}. Specifically, for the Type II model, the upper bound on $\tan\beta$ from $D_s\to\tau\nu_\tau$ is $\tan\beta \leq 50$ with charged Higgs mass $M=600$ GeV \cite{comprehensive}.

In this work we aim to study the charged Higgs phenomenology with large $\tan\beta$ and choose
the Type II Yukawa couplings as the working model 
\begin{eqnarray}
-\mathcal{L} & = & -\cot\beta\frac{m_{u}}{v}\bar{u}_{L}(H+iA)u_{R}\nonumber \\
 &  & +\tan\beta\frac{m_{d}}{v}\bar{d}_{L}(H-iA)d_{R}\nonumber \\
 &  & -\sqrt{2}\cot\beta\frac{m_{u}}{v}V_{ud}^{\dag}\bar{d}_{L}H^{-}u_{R}\nonumber \\
 &  & -\sqrt{2}\tan\beta\frac{m_{d}}{v}V_{ud}\bar{u}_{L}H^{+}d_{R}+{\rm h.c.}.\label{lagrangian}\end{eqnarray}
The vev of SM Higgs is related as $v=\sqrt{v_{1}^{2}+v_{2}^{2}}$. $t\bar{b}H^{-}$ vertex given in Ref.\cite{higgs hunter's guide}
can be written as
 \begin{eqnarray}
g_{H^{-}t\bar{b}}
&=&g_{a}+g_{b}\gamma_{5}.
\end{eqnarray}
Within the type II 2HDM, 
$$g_{a,b}=g({\rm cot}\beta m_{t}\pm {\rm tan}\beta m_{b})/(2\sqrt{2}m_{{\rm W}}).$$ 
In the specific discussions below in Sec. III, we will investigate various 
combinations of $(g_ag_b)$ accordingly not limited in type II 2HDM.

\subsection{Charged Higgs production associated with single top quark at hadron
colliders}

\begin{figure}
\centering \includegraphics[width=0.45\textwidth]{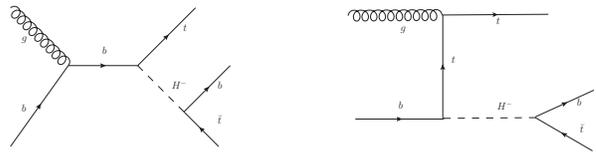}
\caption{Feynman diagrams for $gb\to tH^{-}\to t\bar{t}b$ process. }

\label{fig:com1}
\end{figure}

We begin to consider the following process(Fig.~\ref{fig:com1})
\begin{eqnarray}
\nonumber
g(p_{1})+b(p_{2})&\to& t(p_{3},s_{t})+H^{-}(p_{4})\\
&\to& t(p_{3},s_{t})+b(p_{5})+\bar{t}(p_{6},s_{\bar{t}}),\label{pp2th2ttb}\end{eqnarray}
 where $p_{i}$ denotes the 4-momentum of the corresponding particle.
$s_{t}$($s_{\bar{t}}$) is the top(antitop) quark spin in 4-dimension
and $s_{t}^{2}=s_{\bar{t}}^{2}=-1$, $p_{3}\cdot s_{t}=p_{6}\cdot s_{\bar{t}}=0$.

Under the narrow width approximation of the charged Higgs, i.e., the
charged Higgs is produced on-shell, \begin{equation}
\lim_{\Gamma\to0}\frac{1}{(p_{4}^{2}-M^{2})^{2}+\Gamma^{2}M^{2}}
\longrightarrow\frac{\pi}{\Gamma M}\,\delta(p_{4}^{2}-M^{2}),\end{equation}
 where $\Gamma$ and $M$ respectively denote
the decay width and mass of the charged Higgs boson. The matrix element
squared including top quark spin information for the process (\ref{pp2th2ttb})
can be written as follows 
\begin{eqnarray}
\nonumber
|{\cal M}(s_{t},s_{\bar t})|^{2}=&&|{\cal M}_{gb\to tH^{-}}(s_{t})|^{2}
|{\cal M}_{H^{-}\to b\bar{t}}(s_{\bar{t}})|^{2}\\
&&\times\frac{\pi}{\Gamma M}\delta(p_{4}^{2}-M^{2}),\end{eqnarray}

 where\begin{equation}
|{\cal M}_{gb\to tH^{-}}(s_{t})|^{2}=\frac{g_{s}^{2}}{2N_{c}}\Big\{{\cal A}+
{\cal B}_{1}(p_{1}\cdot s_{t})+{\cal B}_{2}(p_{2}\cdot s_{\bar{t}})\Big\},\label{eqst2}\end{equation}
and \begin{eqnarray}
|{\cal M}_{H^{-}\to b\bar{t}}(s_{\bar{t}})|^{2}&=&(g_{a}^{2}+g_{b}^{2})(M^{2}-m_{b}^{2}
-m_{t}^{2})\nonumber \\
&&-2(g_{a}^{2}-g_{b}^{2})m_{b}m_{t}\nonumber \\
&&-4g_{a}g_{b}m_{t}(p_{5}
\cdot s_{\bar{t}}).\label{eqst1}\end{eqnarray}
 The formula for ${\cal A}$, ${\cal B}_{1}$ and ${\cal B}_{2}$
are listed in the following

\begin{equation}
{\cal A}=(g_{a}^{2}+g_{b}^{2})\, A_{1}\,+\, m_{b}m_{t}(g_{b}^{2}-g_{a}^{2})\, A_{2},\end{equation}
 with 
\begin{eqnarray}
\nonumber
A_{1}&=&\displaystyle{\frac{\hat{s}(p_{1}\cdot p_{3})-m_{b}^{2}(4p_{1}\cdot p_{3}+3p_{2}\cdot p_{3})}
{(\hat{s}-m_{b}^{2})^{2}}}\\
&&+\frac{\hat{s}(p_{1}\cdot p_{3})+m_{t}^{2}(\hat{s}-2p_{2}\cdot
p_{3})}{4(p_{1}\cdot p_{3})^{2}}\nonumber\\
&&-\frac{1}{2(p_{1}\cdot p_{3})(\hat{s}-m_{b}^{2})}\big\{m_{t}^{2}(\hat{s}-2m_{b}^{2})-2(p_{1}\cdot p_{3})m_{b}^{2}\nonumber\\
&&+2(\hat{s}-2p_{2}\cdot
p_{3})(p_{1}\cdot p_{3}+p_{2}\cdot p_{3})\big\},\\
A_{2}&=&\frac{(\hat{s}+2m_{b}^{2})}{(\hat{s}-m_{b}^{2})^{2}}+\frac{m_{t}^{2}-p_{1}\cdot p_{3}}{2(p_{1}
\cdot p_{3})^{2}}\nonumber\\&&-\frac{2p_{1}\cdot p_{3}+4p_{2}\cdot p_{3}-\hat{s}}{2(p_{1}\cdot p_{3})(\hat{s}-m_{b}^{2})},
\end{eqnarray}

and 
\begin{eqnarray}
B_{1}&=&2g_{a}g_{b}m_{t}[\frac{4m_{b}^{2}-\hat{s}}{(\hat{s}-m_{b}^{2})^{2}}+\frac{2p_{2}\cdot p_{3}-\hat{s}}{4(p_{1}
\cdot p_{3})^{2}}+\frac{1}{p_{1}\cdot p_{3}}\nonumber\\&&
-\frac{p_{2}\cdot p_{3}}{(p_{1}\cdot p_{3})(\hat{s}-m_{b}^{2})}],\\
B_{2}&=&2g_{a}g_{b}m_{t}[\frac{3m_{b}^{2}}{(\hat{s}-m_{b}^{2})^{2}}+
\frac{m_{t}^{2}-p_{1}\cdot p_{3}}{2(p_{1}\cdot p_{3})^{2}}\nonumber\\&&+
\frac{\hat{s}-p_{1}\cdot p_{3}-2p_{2}\cdot p_{3}}{(p_{1}\cdot p_{3})(\hat{s}-m_{b}^{2})}].
\end{eqnarray}
 The matrix element squared for
the process $g\bar{b}\to\bar{t}H^{+}\to t\bar{t}\bar{b}$ can be obtained
from the above equations by using CP-invariance.
Obviously, the top quark spin effects are related to 
the product $(g_ag_b)$ and disappear for a pure scalar
or pseudo-scalar charged Higgs boson. In particular, 
for $pp\to tH^-\to t\bar{t}b$ process, 
 this feature can be reflected by the following spin observable
\begin{equation}
\label{double}
<{\cal {O}}_{t}> =2<\bf {{S_{t}}}\cdot  {\bf \hat a}>= {\sigma(\uparrow) - \sigma(\downarrow)\over
\sigma(\uparrow) + \sigma(\downarrow)},\\
\end{equation}
where $\bf{{S_{t}}}$ is top quark spin vector in its rest frame,
and the arrows on the right-hand side refer to the spin state of the top quark  with respect to 
the  quantization axis ${\bf \hat a}$. At LHC, the helicity basis is a better choice, i.e.,
${\bf\hat a}={\bf {\hat{p}}}_t^*$ with the unit 3-momentum ${\bf {\hat{p}}}_t^*$ in the
$tH^-$ center-of-mass frame. 
Similarly, we can also define the spin observable with respect to
antitop quark
\begin{equation}
<{\cal {O}}_{\bar t}> =2<\bf {{S}_{\bar t}}\cdot {\bf \hat b}>,
\end{equation} 
where $\bf{\hat b}$ is the spin quantization axis related to antitop quark. At LHC, we can choose
${\bf\hat b}={\bf {\hat{p}}}_{\bar {t}}^*$ with the unit 3-momentum 
${\bf {\hat{p}}}_{\bar{t}}^*$ in the
charged Higgs rest frame. 

For the subsequent polarized top quark decay 
$$
t({\bf{S_t}}) \to c(p_c)+X,
$$
where $c$ represents a final state particle or jet
and $p_c$ is its momentum,
the corresponding differential distribution is obtained as follows\cite{Brandenburg:2002xr}
\begin{eqnarray}\label{power}
\frac{1}{\Gamma_t}\frac{d\Gamma_t}{d\cos\vartheta}=\frac{1}{2}\left(1+\kappa_c\cos\vartheta\right),
\end{eqnarray}
where $\vartheta$ is the angle between the top quark spin and the moving direction of 
$c$ in top quark rest frame. $\kappa_c$ is the so-called
spin analysing power of the corresponding particle or jet $c$.
For the charged lepton in the semileptonic top quark decay within SM,  
$\kappa_{l^+}=1$ at the tree level.

\section{Numerical Results and discussion}

\label{seciii}

\begin{figure}
\centering \includegraphics[width=0.4\textwidth]{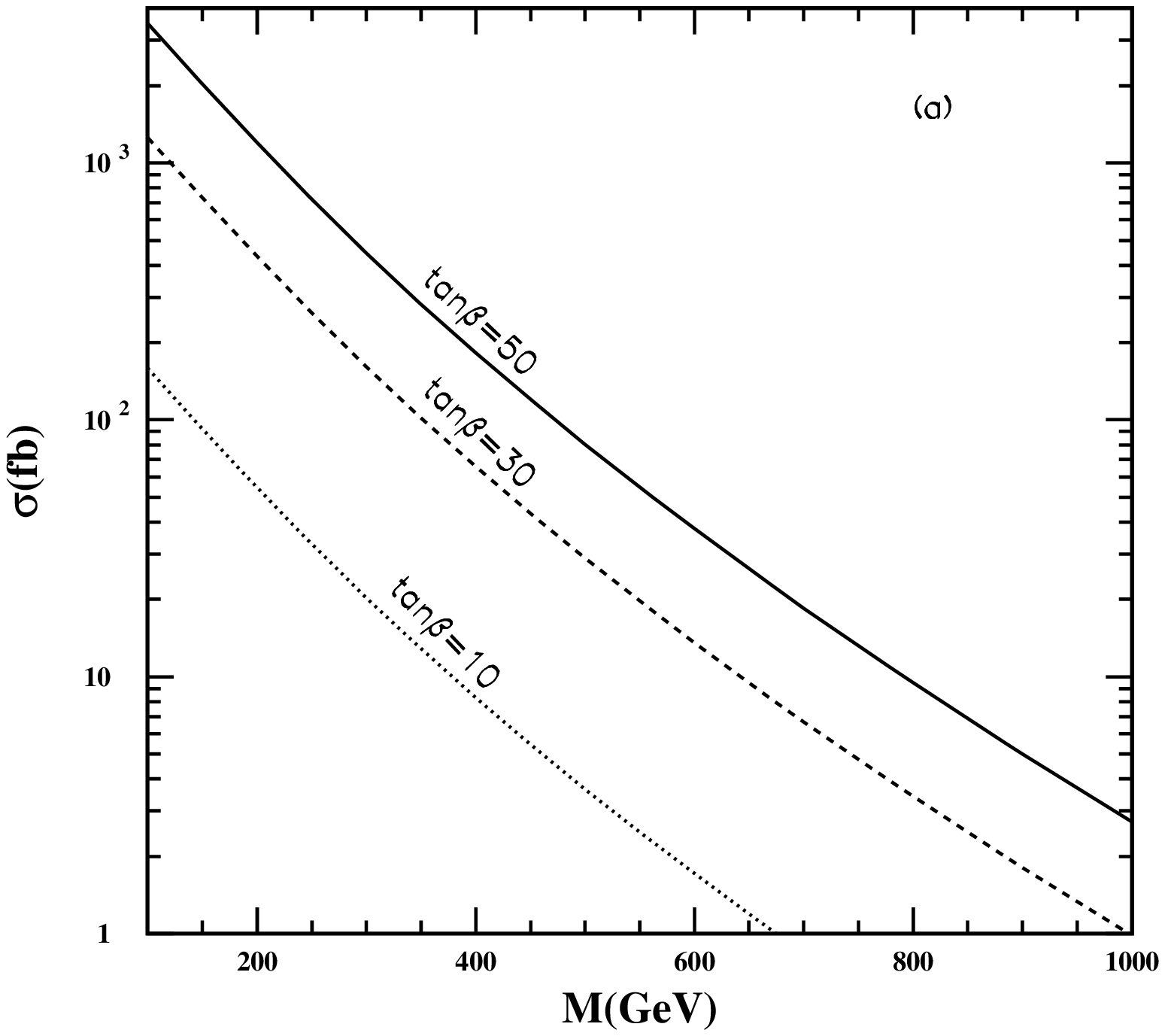} \includegraphics[width=0.4\textwidth]{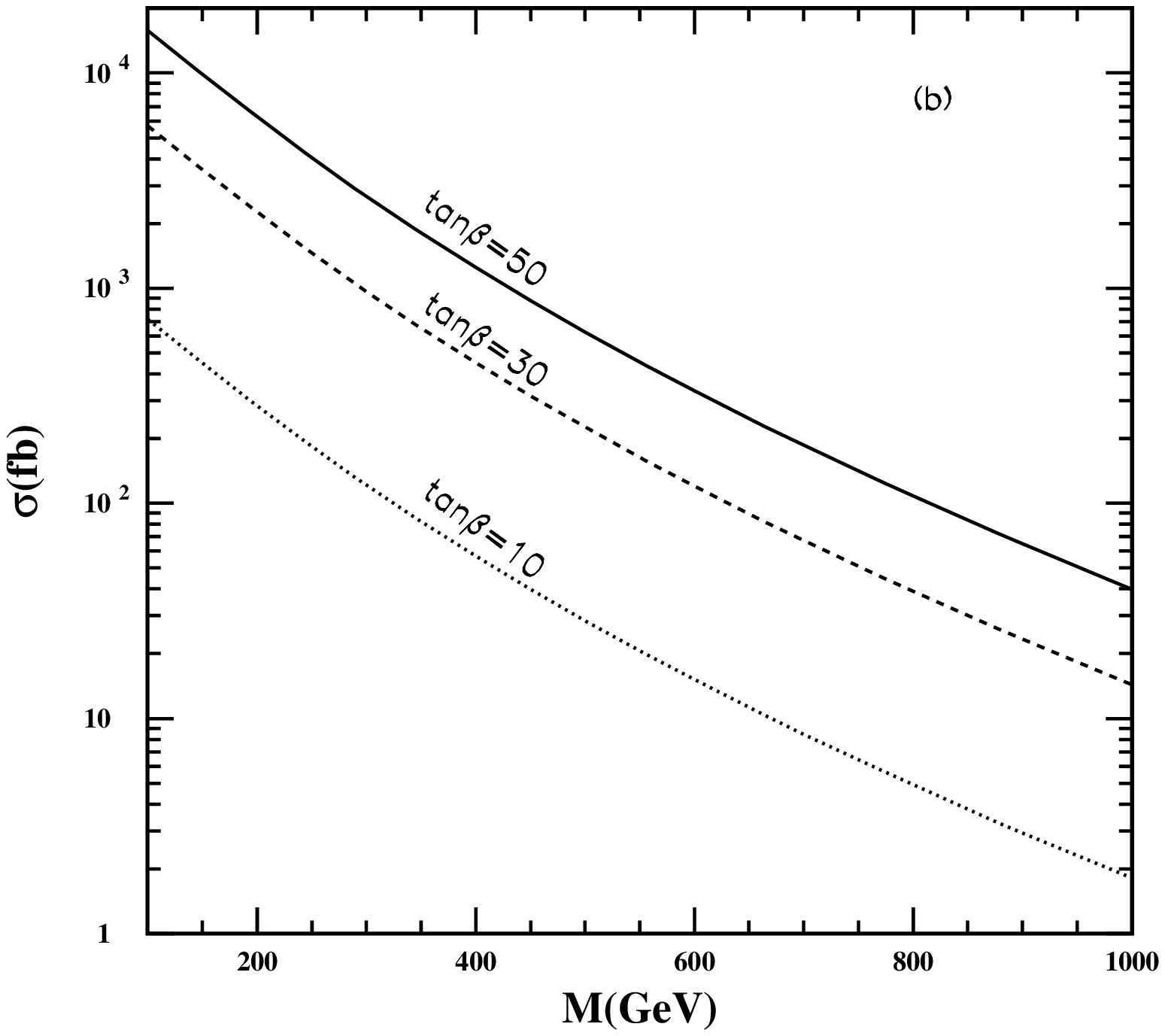}
\caption{The total cross section as a function of $M$ for $pp\to tH^{-}$
process at LHC for (a) 8 TeV and (b) 14 TeV.}

\label{fig:com3}
\end{figure}

For the processes $pp\to tH^{-}$, the total cross section can be
expressed as \begin{equation}
\sigma=\int f_{g}(x_{1})f_{b}(x_{2})\hat{\sigma}_{gb\to tH^{-}}(x_{1}x_{2}s)dx_{1}dx_{2},\end{equation}
 where $f_{g}(x_{1})(f_{b}(x_{2}))$ is the parton distribution function(PDF)
of gluon(quark), $\sqrt{s}$ is the center of mass energy ($c.m.$) of parton-parton collision,
and $\hat{\sigma}$ is the partonic level cross section for $gb\to tH^{-}$
process. In our numerical calculations we set $V_{tb}=1$, $m_{W}=80.399$
GeV, $m_b=4.70$ GeV and $m_{t}=173.1$ GeV. For PDF, we use CTEQ6L1\cite{Pumplin:2002vw}.
In Fig.~\ref{fig:com3}, the total cross sections for the process
$pp\to tH^{-}$ are shown as a function of charged Higgs mass for
$\tan\beta=10$, 30, and 50 in 2HDM at the LHC with 8 TeV and 14 TeV. Obviously, the
$tH^-$ production rate at 14 TeV is much higher than that at 8 TeV.
In the following, we 
investigate the processes
\begin{equation}
pp\to tH^{-}\,\to\, t\bar{t}b\,\to\, bl^{+}\,+\, b\bar{b}jj\,+\,\met,
\label{tlhj}
\end{equation}
\begin{equation}
pp\to tH^{-}\,\to\, t\bar{t}b\,\to\, bjj\,+\, b\bar{b}l^{-}\,+\,\met.
\label{tjhl}\end{equation}
In process (\ref{tlhj}), the top quark produced associated with
$H^{-}$ decays semi-leptonically, and the anti-top quark from charged
Higgs decays hadronically, i.e., $t\to bl^{+}\nu_{l}$ and $\bar{t}\to\bar{b}jj$.
While in process (\ref{tjhl}), ${t}\to{b}jj$ and $\bar{t}\to\bar{b}l^{-}\bar{\nu}_{l}$.
The charged lepton can be used to trigger the event.
The dominant background for the above processes is $pp\to t\bar{t}j$
events.

To be more realistic, the simulation at the detector is performed
by smearing the leptons and jets energies according to the assumption
of the Gaussian resolution parameterization \begin{equation}
\frac{\delta(E)}{E}=\frac{a}{\sqrt{E}}\oplus b,\end{equation}
 where $\delta(E)/E$ is the energy resolution, $a$ is a sampling
term, $b$ is a constant term, and $\oplus$ denotes a sum in quadrature.
We take $a=5\%$, $b=0.55\%$ for leptons and $a=100\%$, $b=5\%$
for jets respectively\cite{atlas0901}.


For our signal process, one top quark which decays hadronically can
be reconstructed from the three jets by demanding $|M_{jjj}-m_{t}|\leq30$GeV,
while to reconstruct another top that is leptonical decay, the 4-momentum
of the neutrino should be known. But the neutrino is an unobservable
particle, so we have to utilize kinematical constraints to reconstruct
its 4-momentum. Its transverse momentum can be obtained by momentum
conservation from the observed particles \begin{equation}
{\bf p}_{\nu T}=-({\bf p}_{lT}+\sum_{j=1}^{5}{\bf p}_{jT}),\end{equation}
 while its longitudinal momentum can not be determined in this way
due to the unknown boost of the partonic $c.m.$ system. Alternatively,
it can be solved with twofold ambiguity through the on shell condition
for the W-boson \begin{equation}
m_{W}^{2}=(p_{\nu}+p_{l})^{2}.\end{equation}
 Furthermore one can remove the ambiguity through the reconstruction
of another top quark. For each possibility we evaluate the invariant
mass \begin{equation}
M_{jl\nu}^{2}=(p_{l}+p_{\nu}+p_{j})^{2},\end{equation}
 where $j$ refers to the any one of the two left jets and pick up
the solution which is closest to the top quark mass. With such a solution,
one can reconstruct the 4-momentum of the neutrino and that of another
top quark.

\begin{figure}
\centering \includegraphics[width=0.40\textwidth]{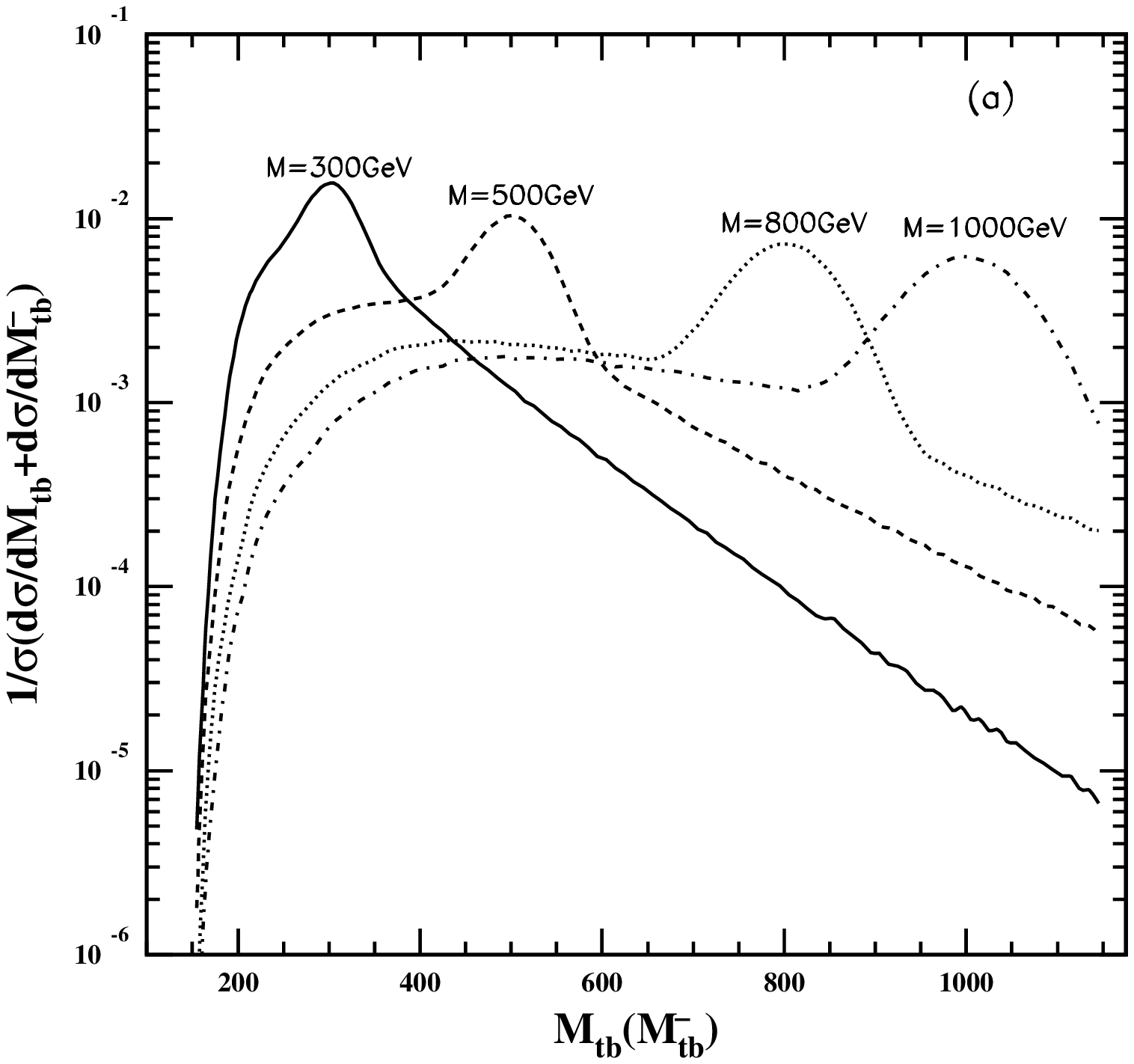} 
\includegraphics[width=0.40\textwidth]{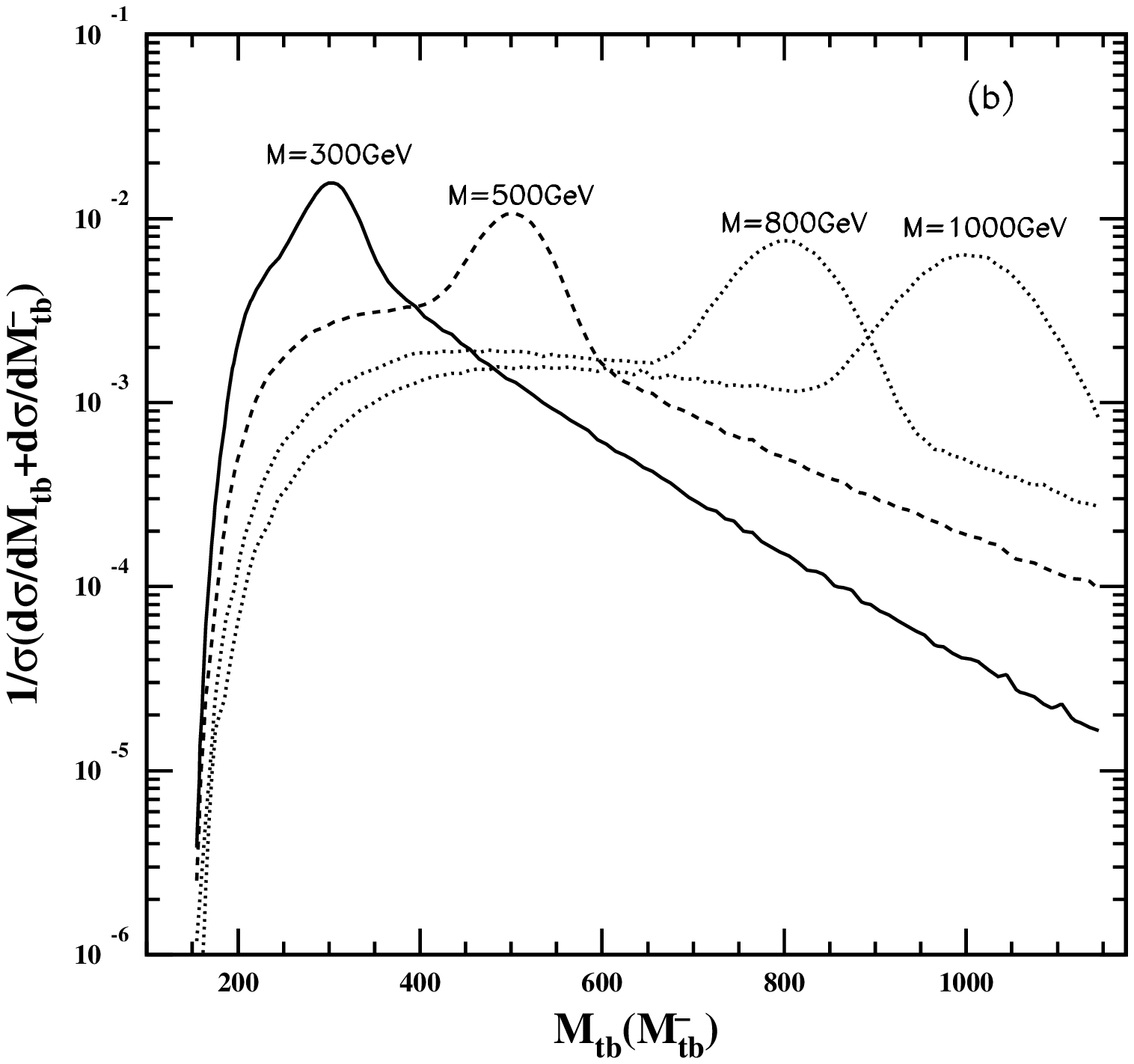}
\caption{The distributions $1/\sigma(d\sigma/dM_{tb}+d\sigma/dM_{\bar{t}b})$ with respect to 
the invariant mass between the reconstructed top (antitop) and the remaining jet 
for the process of $pp\to tH^{-}\to l^{+}+5jets+\met$ after cut I at LHC 
for (a) 8 TeV and (b) 14 TeV.}

\label{fig:mtb}
\end{figure}

In our following numerical calculations, we apply the basic acceptance
cuts(refered to as cut I)
\begin{eqnarray}
 &  & p_{lT}>20~{\rm GeV},~~~~p_{jT}>20~{\rm GeV},~~~~\met>20~{\rm GeV},\nonumber \\
 &  & |\eta_{l}|<2.5,~~~~|\eta_{j}|<2.5,~~~~\Delta R_{jj(lj)}>0.4,\nonumber \\
 &  & |M_{j_{a}l\nu}-m_{t}|\leq30~{\rm GeV},~~~~|M_{j_{b}j_{c}j_{d}}-m_{t}|\leq30~{\rm GeV},\nonumber\\
&&|M_{j_{b}j_{c}}-m_{W}|<10~{\rm GeV}.
\label{cut1}\end{eqnarray}
Once the events were fully reconstructed after smearing and including the cut I,
we can further reconstruct the invariant mass between the reconstructed top (antitop) 
and the remaining jet. We display the distributions  
$1/\sigma(d\sigma/dM_{tb}+d\sigma/dM_{\bar{t}b})$ in Fig.~\ref{fig:mtb},
where the resonance peak can easily be seen for different charged Higgs masses.
Due to the existence of the resonance peak, we further employ another cut(refered to as cut II)
\begin{itemize}
\item Cut II: $|M_{jj_{b}j_{c}j_{d}}-M|\leq10\%\, M$
or $|M_{jj_{a}l\nu}-M|\leq10\%\, M$.
\end{itemize}
Comparing our signal process with the dominant background process $pp\to t\bar{t}j$,
it is easy to notice that, after reconstructing the top and antitop quarks,
the remaining jet is a b-jet for our signal while it is  a light jet for the background 
with  large probability.
Therefore to further purify the signal, we adopt the following cut(refered to as cut III)
\begin{itemize}
\item Cut III: We demand the remaining jet that cannot be used to reconstruct
top quarks to be a b-jet.
\end{itemize}
The $b$-tagging efficiency is assumed to be 60\% while the miss-tagging efficiency of a light jet as
a $b$ jet is taken as transverse momentum dependent \cite{atlas0901}:
\begin{equation}
\epsilon_l =\left\{ \begin{array}{ll}
\displaystyle{\frac{1}{150}}, &P_T< 100\, {\rm GeV},\\
\displaystyle{\frac{1}{450}}\,[\frac{P_T}{25\,{\rm GeV}}-1], &100 \,{\rm GeV}\leq P_T<250\, {\rm GeV},\\
\displaystyle{\frac{1}{50}}, & P_T\geq 250\, {\rm GeV}.
\end{array} \right.
\end{equation}

The cross sections for the signal processes (\ref{tlhj}) and (\ref{tjhl})
with different charged Higgs mass after each cuts at
LHC 8 and 14 TeV are respectively listed in Table.~\ref{tab:csa}
and \ref{tab:csb}.
The dominant SM background related to the signal is
$pp\to t\bar{t}j\to l^\pm+5jets+\met$
process. We employ  MadEvent\cite{Madevent}
to simulate the background processes.  The other SM background processes, e.g., $Wjjjjj$,
$WWjjj$ and $WZjjj$, etc. are dramatically reduced by the acceptance cuts we adopt
and therefore we neglected them here.
Supposing the integral luminosity to be 20 $fb^{-1}$ at
$\sqrt{s}=8$ TeV, one can notice that it is difficult for the charged
Higgs associated with a top quark to be detected when its
mass is above 500 GeV. While with $300 fb^{-1}$ integral luminosity at 14 TeV, the $tH^-$ production is easier
to be observed. Detailed analysis
shows that for  $pp\to tH^- \to t\bar{t}b\to l^+ (or~ l^-)+bb\bar{b}jj+\met$ process at 14 TeV ,
the significance of signal to background can also be above three sigma
for the charged Higgs mass $M\leq 1$ TeV. Therefore, in the following, 
we will focus on the $tH^-$ production 
at  $\sqrt{s}=14$ TeV.

From Eqs. (\ref{eqst2}) and (\ref{eqst1}), one can notice  that top
quark spin effect is related to the product $(g_{a}g_{b})$.
Using the same method as in Ref.\cite{bbsu}, we find that this kind of top quark spin effects
can be translated into the angular distributions of the charged lepton.
Corresponding to the process (\ref{tlhj}) and (\ref{tjhl}), we obtain 
two kinds of angular distributions
\begin{eqnarray}
&&\frac{1}{\sigma}\,\frac{d\sigma}{d\cos\theta^{*}}\,=\,\frac{1}{2}\,[1\,+\, A_{FB}\cos\theta^{*}],\nonumber\\
&  & \frac{1}{\sigma}\,\frac{d\sigma}{d\cos\bar{\theta}^{*}}\,=\,\frac{1}{2}\,[1\,+\,
\bar{A}_{FB}\cos\bar{\theta}^{*}],
\label{eqfbeq}\end{eqnarray}
 with
\begin{equation}
\cos\theta^{*}={{\bf\hat{p}}_{l^{+}}^{*}\cdot{\bf \hat{p}}_{t}^{*}},
~~~~~\cos\bar{\theta}^{*}={{\bf \hat{p}}_{l^{-}}^{*}\cdot{\bf \hat{p}}_{\bar{t}}^{*}},
\end{equation}
where ${\bf \hat{p}}_{l^{+}}^{*}$ is the unit 3-momentum of charged lepton in
the top quark rest frame, and ${\bf \hat{p}}_{l^{-}}^{*}$
is the unit 3-momentum of charged lepton in the anti-top quark rest frame.
Without smearing effect and any acceptance cuts,
$A_{FB}$($\bar{A}_{FB}$) can be related to 
the top quark spin observables ${\cal{O}}_t$(${\cal{O}}_{\bar{t}}$), i.e.,
\begin{eqnarray}
&&A_{FB}=\kappa_c <{\cal O}_t>=2\kappa_c<\bf {{S_{t}}}\cdot  {\bf \hat p}_t^*>,\\
&&\bar{A}_{FB}=\kappa_c <{\cal O}_{\bar{t}}>=2\kappa_c<\bf {{S}_{\bar t}}\cdot {\bf \hat p}_{\bar{t}}^*>.
\end{eqnarray}
Obviously for the charged lepton in semileptonic top quark decay, 
$A_{FB}=<{\cal O}_t>$ and $\bar{A}_{FB}= <{\cal O}_{\bar{t}}>$ before smearing effect and cuts.

\begin{figure}
\centering \includegraphics[width=0.40\textwidth]{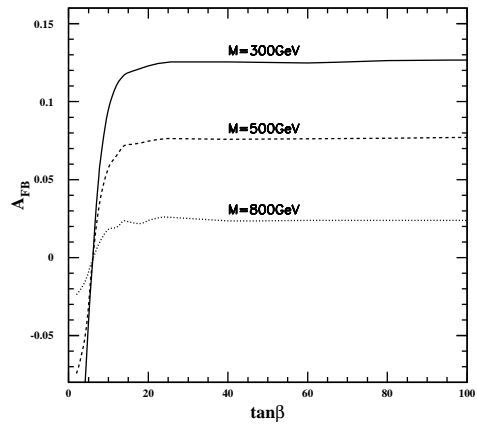} 
\caption{The $A_{FB}$ before any acceptance cuts with respect to
$\tan\beta$ at LHC 14TeV.}
\label{fig:afb}
\end{figure}

\begin{figure}
\centering \includegraphics[width=0.40\textwidth]{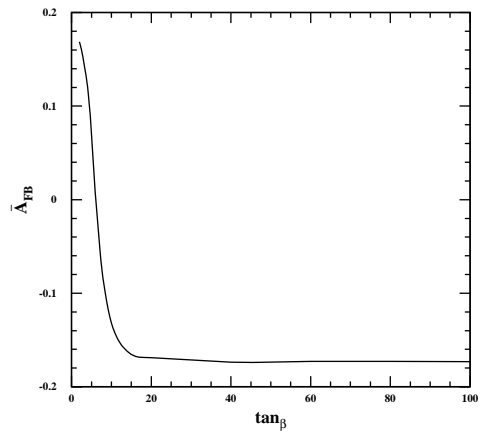} 
\caption{The $\bar{A}_{FB}$ before any acceptance cuts with respect to
$\tan\beta$.}
\label{fig:abarfb}
\end{figure}

\begin{figure}
\centering \includegraphics[width=0.30\textwidth]{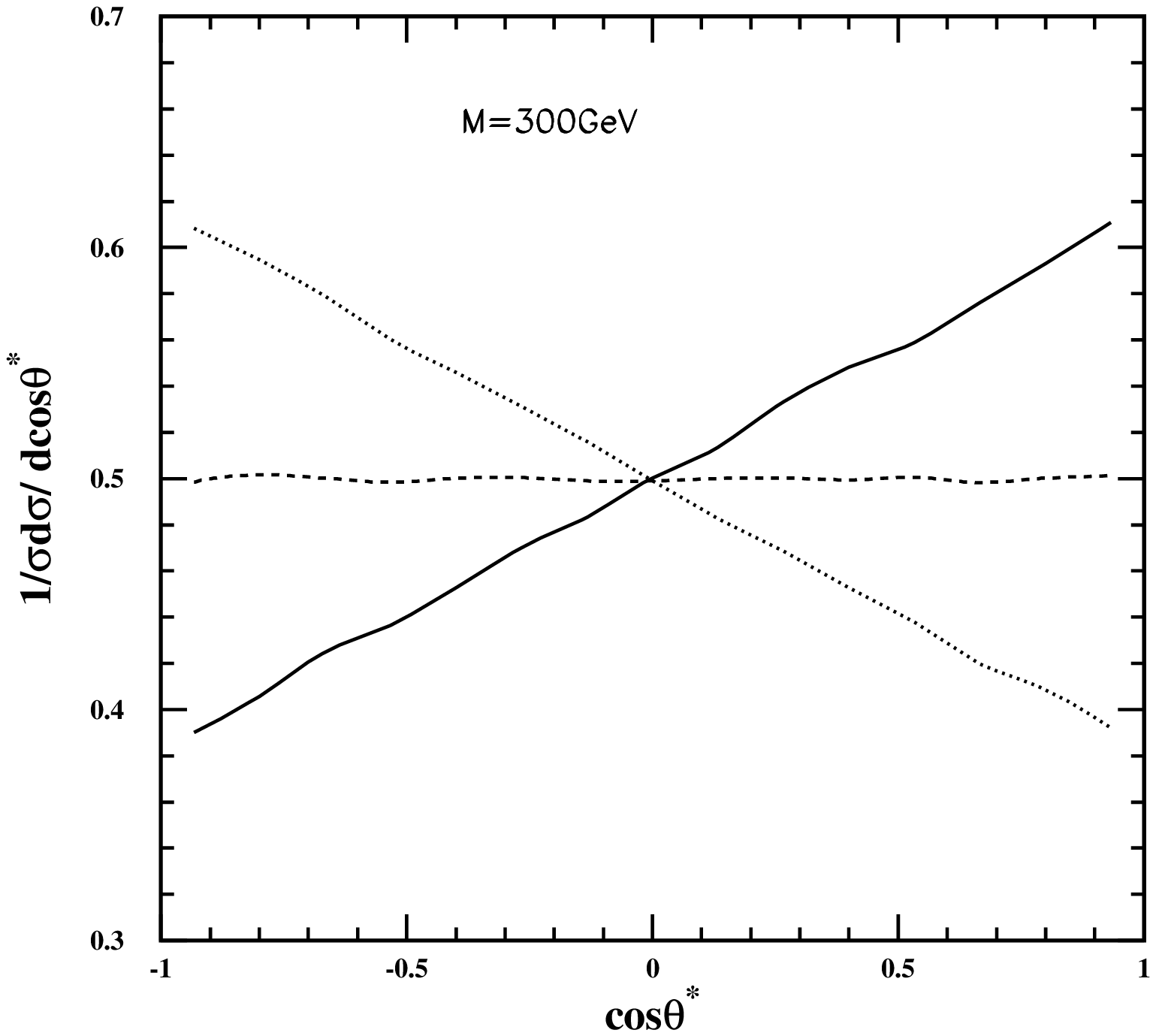}
\includegraphics[width=0.30\textwidth]{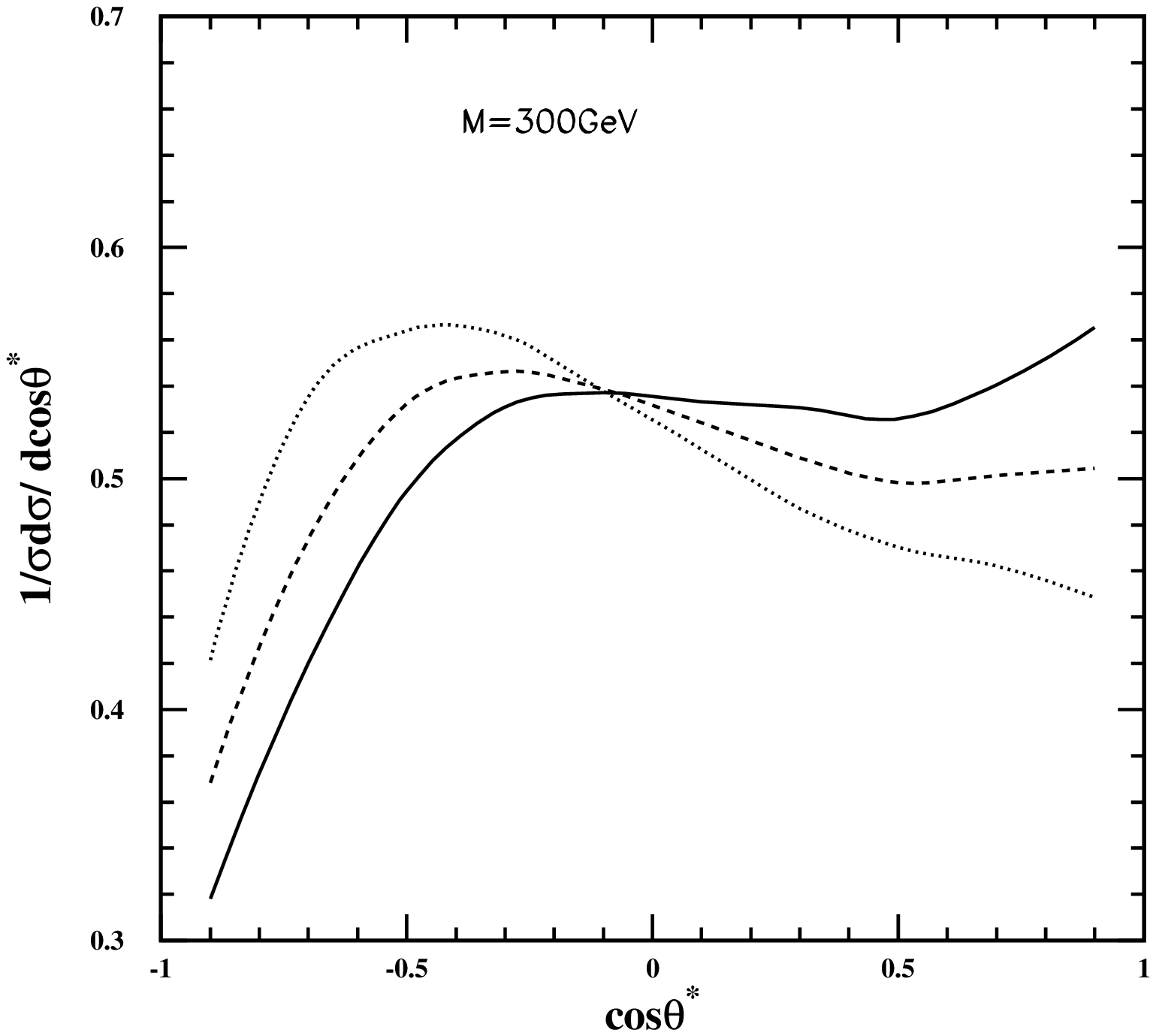}
\includegraphics[width=0.30\textwidth]{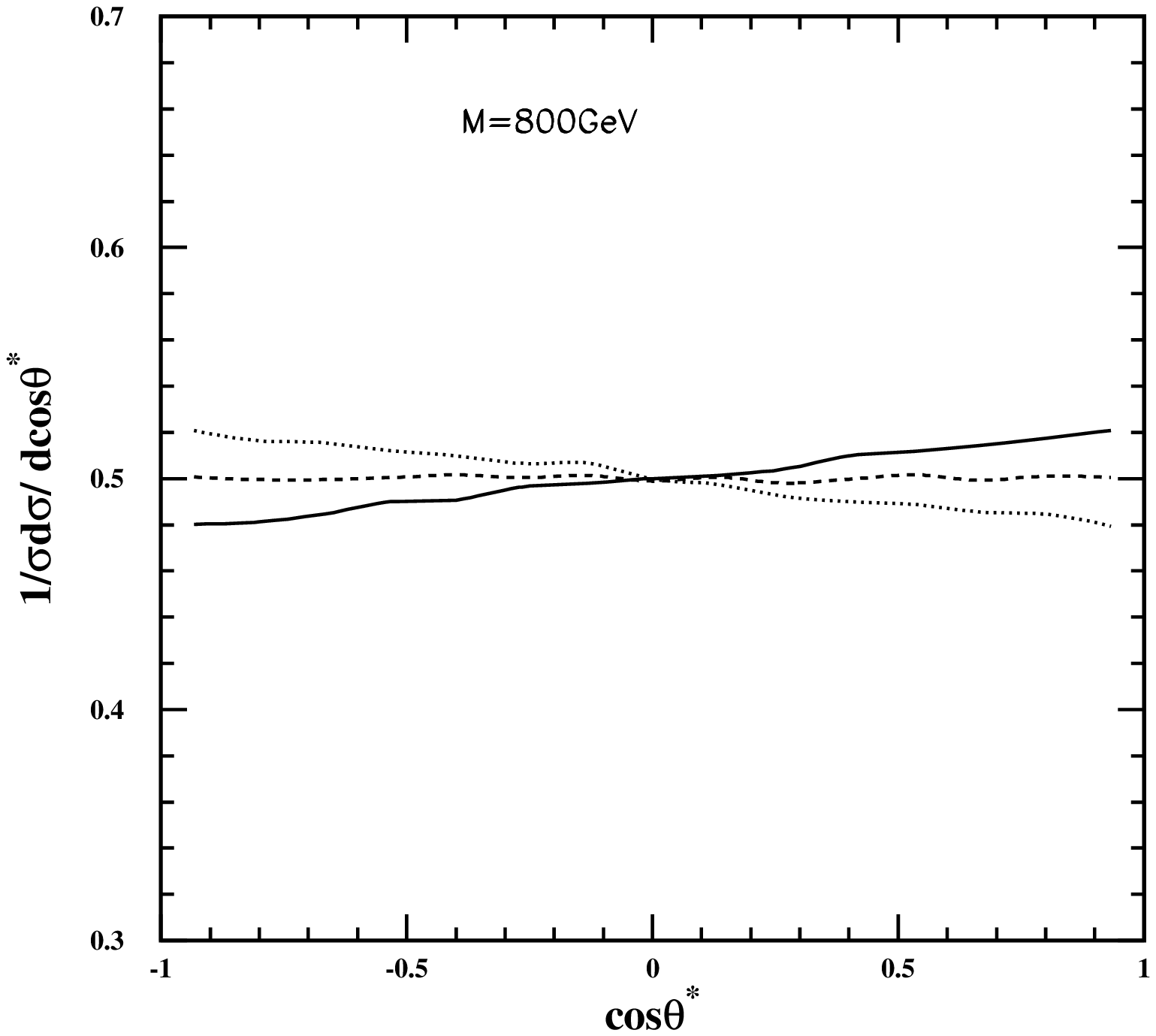}
\includegraphics[width=0.30\textwidth]{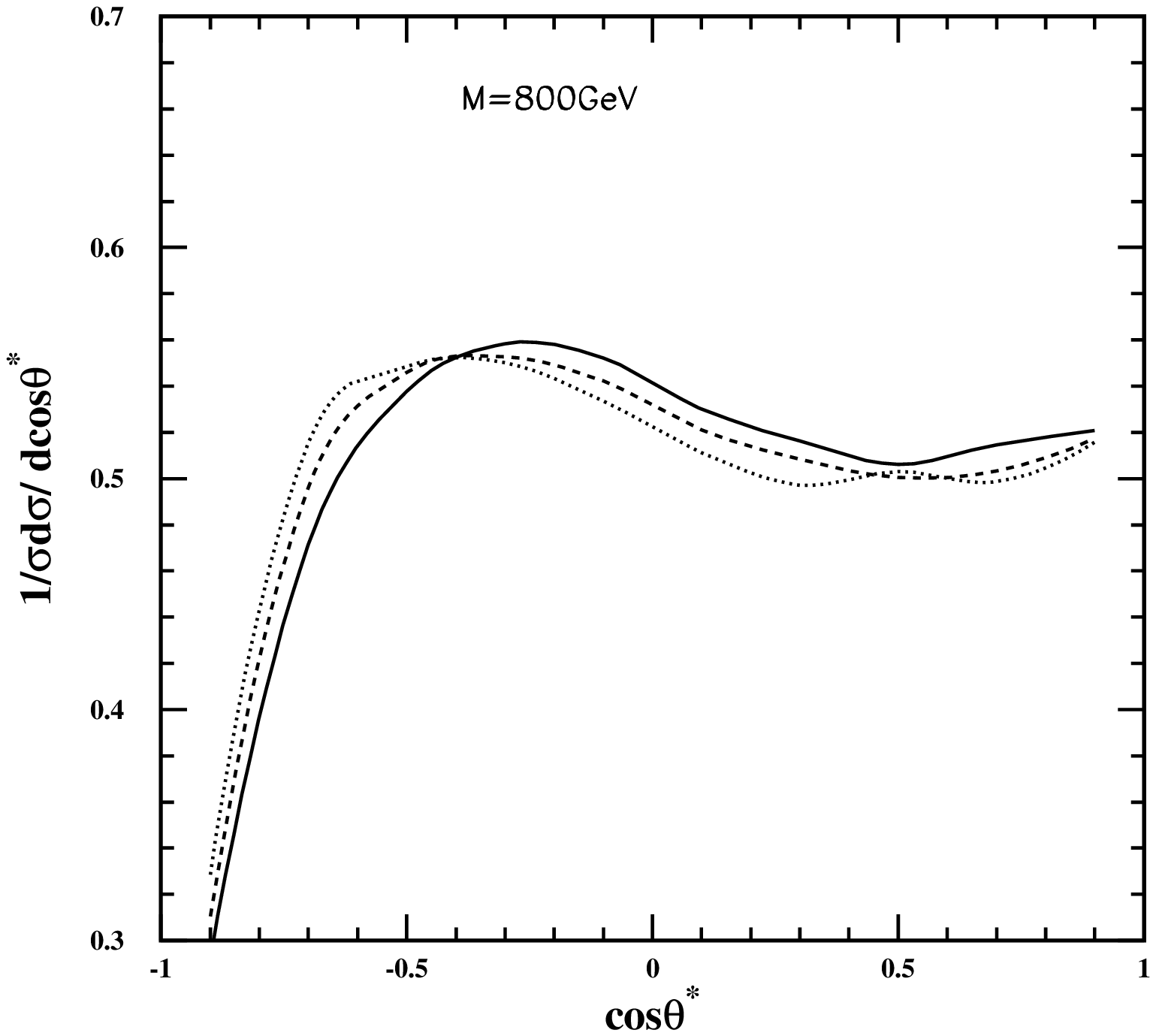}
\caption{The angle distribution of the
charged lepton for $M=$ 300,~800 GeV with nocut and all cuts
at $\sqrt{s}=14$ TeV respectively
for the process of $pp\to tH^{-}\to l^{+}+5jets+\met$.The solid line
represents $(g_{a}g_{b})>0$. The dashed line represents
$(g_{a}g_{b})=0$. The dotted line represents $(g_{a}g_{b})<0$.}

\label{fig:cos-1}
\end{figure}

\begin{figure}
\centering \includegraphics[width=0.30\textwidth]{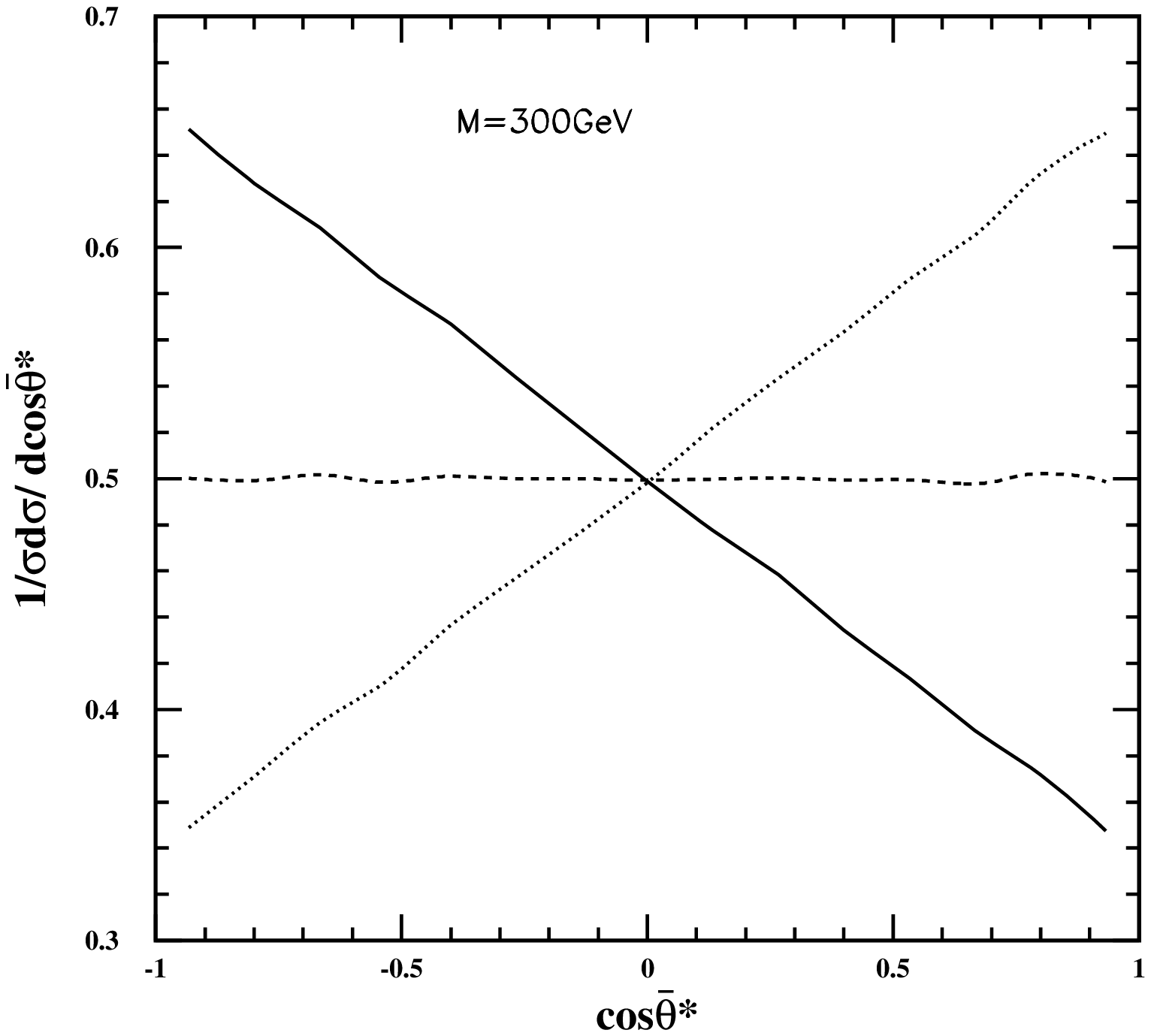}
\includegraphics[width=0.30\textwidth]{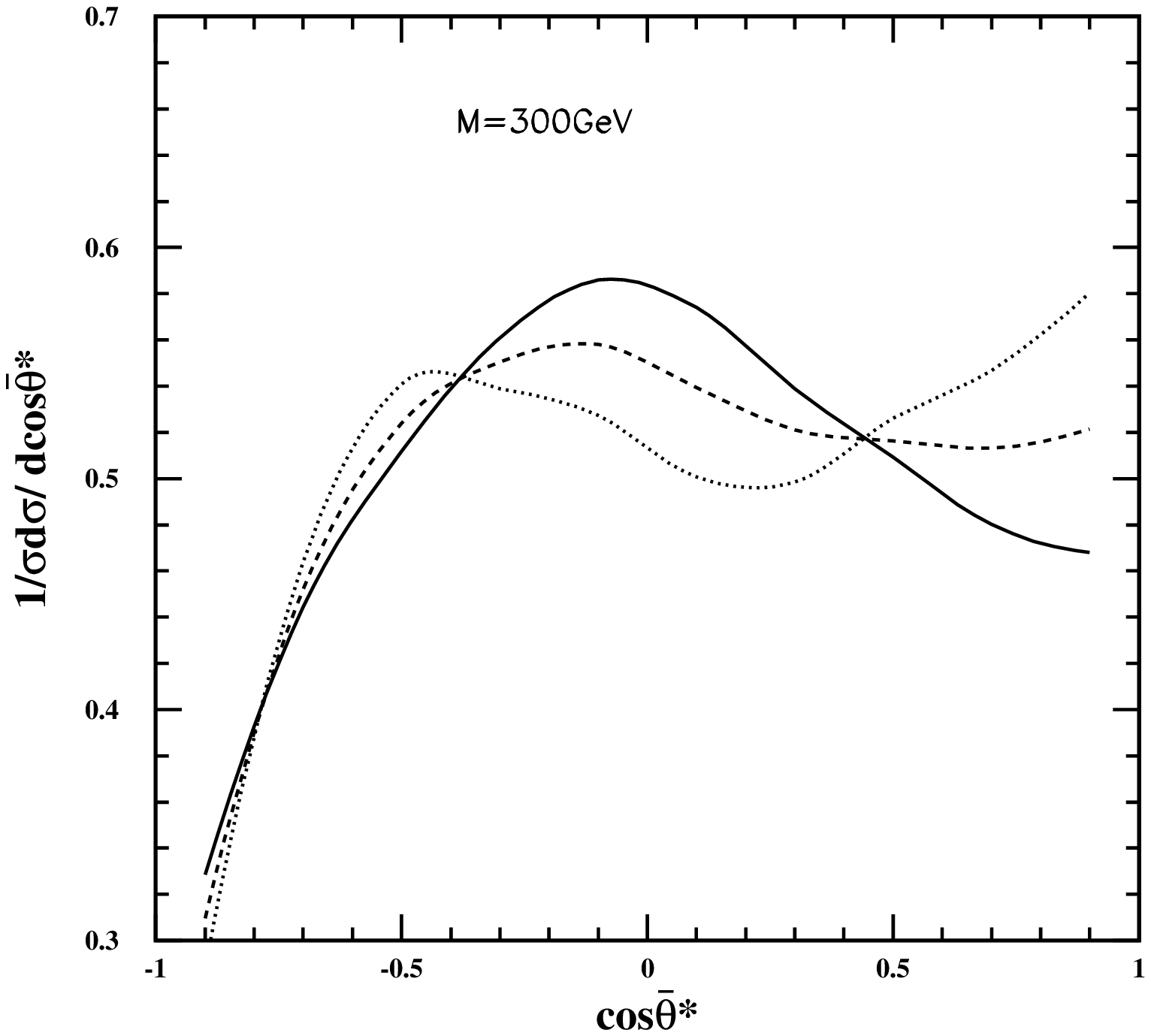}
\includegraphics[width=0.30\textwidth]{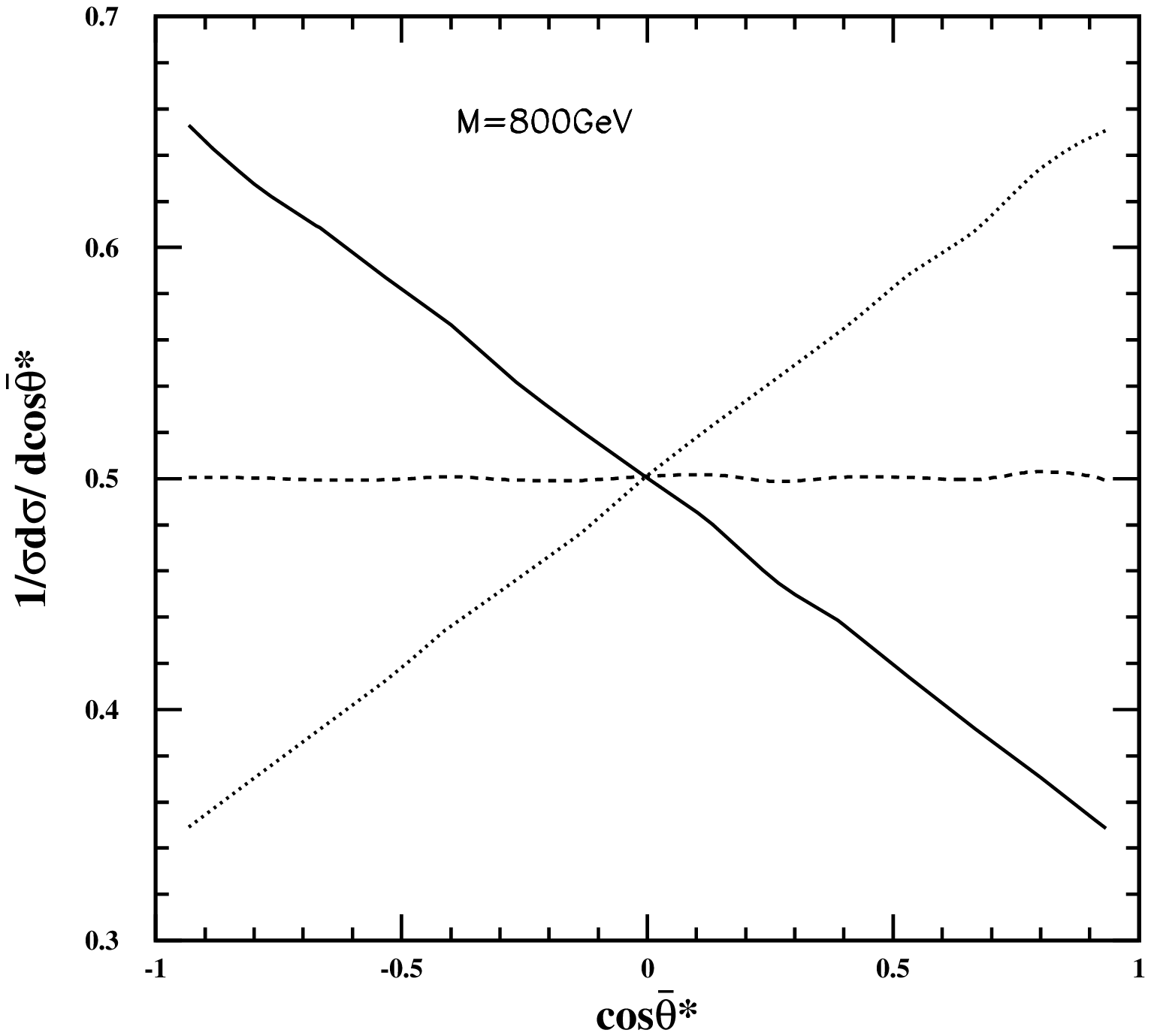}
\includegraphics[width=0.30\textwidth]{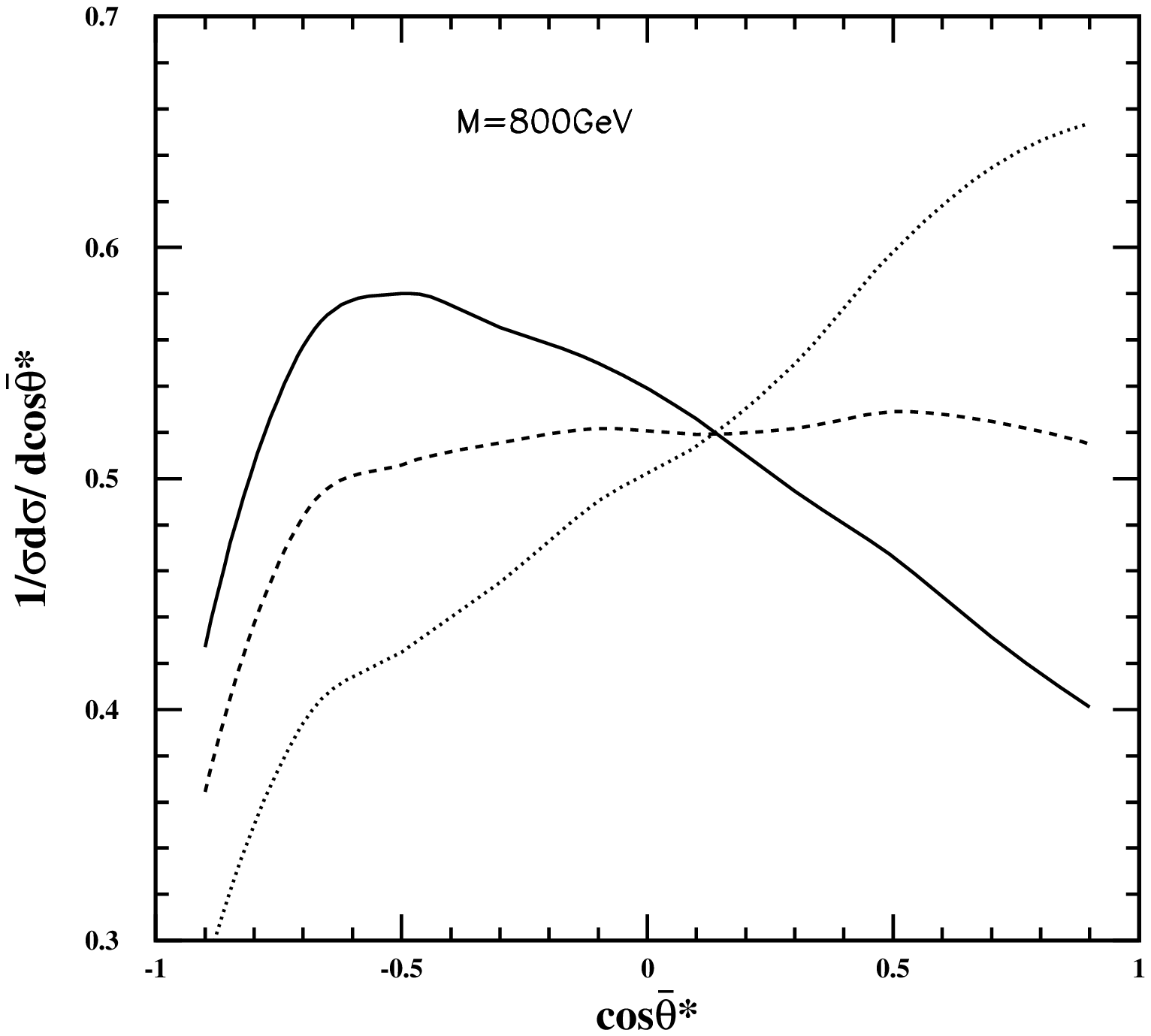}
\caption{The angle distribution  of charged lepton for
$M=$ 300,~800 GeV with nocut and all cuts at
$\sqrt{s}=14$ TeV respectively
for the process of $pp\to tH^{-}\to l^{-}+5jets+\met$.The solid line
represents $ (g_{a}g_{b})>0$. The  dashed line represents
$ (g_{a}g_{b})=0$. The dotted line represents $(g_{a}g_{b})<0$.}

\label{fig:cos-2}
\end{figure}

According to Eq.(\ref{eqfbeq}), $A_{FB}$ and $\bar{A}_{FB}$ can also be determined as follows
\begin{eqnarray}
&&A_{FB}=\frac{\sigma(\cos\theta^{*}>0)-\sigma(\cos\theta^{*}<0)}
{\sigma(\cos\theta^{*}>0)+\sigma(\cos\theta^{*}<0)},\nonumber\\
&&\bar{A}_{FB}=\frac{\sigma(\cos\bar{\theta}^{*}>0)-\sigma(\cos\bar{\theta}^{*}<0)}
{\sigma(\cos\bar{\theta^{*}}>0)+\sigma(\cos\bar{\theta}^{*}<0)}.
\end{eqnarray}
As discussed above, these observables can be used to discriminate different 
$Htb$ interactions.  
Within the framework of type II 2HDM, the form of the Yukawa coupling is fixed with
only one free parameter $\tan \beta$. As displayed in Fig. \ref{fig:afb}, we calculate the 
values of $A_{FB}$ before any acceptance cuts with respect to
$\tan\beta$ for M=300, 500 and 800 GeV at LHC 14TeV. The results for $\bar{A}_{FB}$ before 
any acceptance cuts are
also shown in Fig. \ref{fig:abarfb}. $\bar{A}_{FB}$ is related to the charged Higgs boson decay and we find 
it does not depend on the charged Higgs mass. Combining the results of the $tH^-$ production cross section 
together with the forward-backward asymmetry  $A_{FB}$  and $\bar{A}_{FB}$, one can abstract the useful information of
$\tan\beta$ related to the $Htb$ coupling.
 Next, we  extend our discussions beyond 2HDM and
investigate the $Htb$ coupling in a 
general model where the Yukawa coupling of the charged Higgs bosons to fermions is
a free parameter and so one can regard 
the scalar and pseudo-scalar parts of the Yukawa coupling as
completely independent and free parameters.
In the following, as an example,  we choose  $\tan\beta=30$ and
investigate the charged lepton angular distributions for three different
combinations of $(g_{a} g_{b})$:
\begin{itemize}
\item $(g_ag_b)>0$,   e.g.,\\
$g_{a}=\pm g({\rm cot}\beta m_{t}+{\rm tan}\beta m_{b})/(2\sqrt{2}m_{{\rm W}})$, \\
$g_{b}=\pm g(\cot\beta m_{t}-\tan\beta m_{b})/(2\sqrt{2}m_{{\rm W}})$.
\item $(g_ag_b)=0$,  e.g.,\\ 
$g_{a}=0$, $g_{b}=g(\cot\beta m_{t}-\tan\beta m_{b})/(2\sqrt{2}m_{{\rm W}})$ or
$g_{a}=g({\rm cot}\beta m_{t}+{\rm tan}\beta m_{b})/(2\sqrt{2}m_{{\rm W}})$, $g_b=0$.
\item $(g_ag_b)<0$, e.g.,\\
$g_{a}=\pm g({\rm cot}\beta m_{t}+{\rm tan}\beta m_{b})/(2\sqrt{2}m_{{\rm W}})$,\\
$g_{b}=\mp g(\cot\beta m_{t}-\tan\beta m_{b})/(2\sqrt{2}m_{{\rm W}})$.
\end{itemize}
The charged lepton angular distributions with respect to
$\cos\theta^*$ and $\cos\bar{\theta}^*$  before and after all cuts are respectively
shown in Figs. \ref{fig:cos-1} and \ref{fig:cos-2}. The related
predictions for $A_{FB}$ and $\bar{A}_{FB}$ are listed in Table \ref{tab:asy}.
Due to the fact that the contribution from the $s$-channel(Fig.\ref{fig:com1}(a))
decreases as the charged Higgs mass increases, before all the acceptance cuts,
the $\cos\theta^*$ distribution and
the related results for $A_{FB}$ which are related to $tH^-$ production
 also depend on $M$; while the $\cos\bar{\theta}^*$ distribution and
the related results for $\bar{A}_{FB}$ which are related to the charged Higgs decay
do not depend on $M$.
The $\cos\theta^*=-1$($\cos\bar{\theta}^*=-1$) region
corresponds to lepton that is emitted into the hemisphere
opposite to the top(antitop) direction
of flight in the $tH^-$ $c.m.$ frame. These leptons are
therefore less energetic on average and thus more strongly
affected by the acceptance cuts than those in the remaining region\cite{breuthersi}. Therefore
the presence of the acceptance cuts severely distort these distributions
in the vicinity of $\cos\theta^*=-1$($\cos\bar{\theta}^*=-1$) region
as shown in Figs. \ref{fig:cos-1} and \ref{fig:cos-2}. Therefore for $A_{FB}$ and $\bar{A}_{FB}$
after all acceptance cuts,
we choose $\cos\theta$ ranging from $-0.5$ to $0.5$. It seems that after the acceptance cuts,
the angular distribution with respect to $\cos\theta^*$($\cos\bar{\theta}^*$) and $A_{FB}$($\bar{A}_{FB}$) is
more helpful to investigate the $Htb$ interaction for light(heavy) charged Higgs production associated with
top quark at LHC.

%

%
\begin{table}
\begin{centering}
\begin{tabular}{|c||c|c|c|c|}
\hline
~Signal & \multicolumn{4}{c|}{$\sigma(pp\to tH^{-}\to t\bar{t}b\to l^{\pm}+5jets+\met)$ (fb)}\tabularnewline
\hline
$M$(TeV) & 0.3 & 0.5 & 0.8 & 1.0\tabularnewline
\hline
No cuts  &  45.18 &  8.62 &  1.02 &  0.30 \tabularnewline
\hline
Cut I   &   11.72 &  2.20 &  0.25 &  0.07  \tabularnewline
\hline
+Cut II &   9.59 &  1.73 &  0.20 &  0.05  \tabularnewline
\hline
+Cut III &    5.76 &  1.04 &  0.12 &  0.03\tabularnewline
\hline  \hline
Background & \multicolumn{4}{c|}{$\sigma(pp\to t\bar{t}j\to l^{\pm}+5jets+\met)$ (fb)} \tabularnewline
\hline
Cuts I+II+III &  10.25 &  3.85 &  0.97 &  0.46\tabularnewline
\hline  \hline
$S/B$         &  0.56 &  0.27 &  0.12 &  0.07    \tabularnewline
\hline
$S/\sqrt{B}$ &   8.05 &  2.37 &  0.54 &  0.20\tabularnewline
\hline
\end{tabular}\caption{The cross section of the signal process $pp\to tH^{-}\to l^{\pm}+5jets+\met$
and the background process of $pp\to t\bar{t}j\to l^{\pm}+5jets+\met$
at $\sqrt{s}=8$ TeV after each cut.}

\label{tab:csa}
\par\end{centering}

\centering{}
\end{table}

\begin{table}
\begin{centering}
\begin{tabular}{|c||c|c|c|c|}
\hline
~Signal & \multicolumn{4}{c|}{$\sigma(pp\to tH^{-}\to t\bar{t}b\to l^{\pm}+5jets+\met)$ (fb)}
\tabularnewline
\hline
$M$(TeV) & 0.3 & 0.5 & 0.8 & 1.0 \tabularnewline
\hline
No cuts  &   262.82  & 65.96  & 11.56  &  4.28 \tabularnewline
\hline
Cut I    &    65.79  & 16.41  &  2.75  &  0.95\tabularnewline
\hline
+Cut II  &   54.12  & 13.00  &  2.20  &  0.77\tabularnewline
\hline
+Cut III  &   32.47  &  7.80  &  1.32  &  0.46\tabularnewline
\hline  \hline
Background & \multicolumn{4}{c|}{$\sigma(pp\to t\bar{t}j\to l^{\pm}+5jets+\met)$ (fb)} \tabularnewline
\hline
Cuts I+II+III &   43.54  & 18.97  &  5.95  &  3.22 \tabularnewline
\hline  \hline
$S/B$        &     0.75 &  0.41 &  0.22 &   0.14\tabularnewline
\hline
$S/\sqrt{B}$ &   85.23 & 31.02 &  9.37 &   4.44\tabularnewline
\hline
\end{tabular}\caption{The cross section of the signal process $pp\to tH^{-}\to l^{\pm}+5jets+\met$
and the background process of $pp\to t\bar{t}j\to l^{\pm}+5jets+\met$
at $\sqrt{s}=14$ TeV after each cut.}

\label{tab:csb}
\par\end{centering}

\centering{}
\end{table}


\begin{table}
\begin{centering}\scalebox{0.9}{$
\begin{tabular}{|c||c|c|c|c||c|c|c|c|}
\hline
 & \multicolumn{4}{c||}{$A_{FB}$}
 & \multicolumn{4}{c|} {$\bar{A}_{FB}$}\tabularnewline
\hline \hline
 \multicolumn{9}{|c|}{without cuts and smearing effect} \tabularnewline
\hline \hline
$M$(TeV) & 0.3 &0.5 & 0.8 & 1.0 & 0.3 & 0.5 & 0.8 & 1.0 \tabularnewline
\hline
$(g_ag_b)>0$  & 0.124  & 0.075  & 0.023  & -0.003  & -0.173  & -0.173 & -0.172  & -0.173\tabularnewline
\hline
$(g_ag_b)=0$  & 0.0 & 0.0  & 0.0  & 0.0  & 0.0  & 0.0  & 0.0  & 0.0   \tabularnewline
\hline
$(g_ag_b)<0$  & -0.125  & -0.076 & -0.024 & 0.001  & 0.172  & 0.173 & 0.174  & 0.173\tabularnewline
\hline  \hline
 \multicolumn{9}{|c|}{with cuts and smearing effect} \tabularnewline
\hline \hline
$(g_ag_b)>0$   & 0.002 & -0.015  & -0.031  & -0.041  & -0.014  & -0.050 & -0.054  & -0.061\tabularnewline
\hline
$(g_ag_b)=0$   & -0.028  & -0.030 & -0.033  & -0.041  & -0.022  & 0.007 & 0.006 & -0.006\tabularnewline
\hline
$(g_ag_b)<0$   & -0.056  & -0.048 & -0.037 & -0.040  & -0.033  & 0.081& 0.077  & 0.065\tabularnewline
\hline
\end{tabular}$}\caption{The forward-backward asymmetry $A_{FB}$($\bar{A}_{FB}$)
for $pp\to tH^{-}\to l^{+}(l^-)+5jets+\met$ at LHC $\sqrt{s}=14$ TeV before
and after all cuts.}

\label{tab:asy}
\par\end{centering}

\centering{}
\end{table}


\section{Summary}

The observation of charged Higgs would be an unambiguous signal for the existence
of new physics beyond SM. Therefore it is important to study the related
phenomena both at theory and experiments. In this paper, we study
the $tH^{-}$ associated production via $pp\to tH^{-}\to t\bar{t}b\to l^\pm +bb\bar{b}jj+\met$
process at LHC. It is found that with 300 $fb^{-1}$ integral luminosity at
 $\sqrt{s}=14$ TeV, the signal
can be distinguished from the backgrounds for the charged
Higgs mass up to  1 TeV or  even larger. If the $tH^-$ production
is observed at LHC, one of the key questions is to identify
the $Htb$ interaction. For this aim, we investigate the angular
distributions of the charged leptons and the related forward-backward
asymmetry induced by top quark spin. It is found that these distributions and
observables are sensitive to the product $(g_{a}g_{b})$, so that they can be
used to identify the $Htb$ interaction.
Though further studies
are still necessary both at theory and experiments, the $Htb$ interaction can be studied
by the help of the charged
lepton angular distribution and the related forward-backward asymmetry
in the charged Higgs and top quark associated production at LHC. 
Our analyses are helpful to discriminate the $Htb$ interaction in the 
2HDM or the other new physics models including the charged Higgs.

\section*{Acknowledgements}

This work was supported in part by the National Science
Foundation of China (NSFC), Natural Science Foundation
of Shandong Province (JQ200902) (Z.S.), NSFC (No.11175251 and 11205023) (S.Y.) and NSC (Y.Z.). The authors would like to
thank Dr. S. Bao, Profs. S. Li and X. He for
their helpful discussions.



\begin{thebibliography}{27}
\bibitem{Higgs_ATLAS} G.~Aad \textit{et al.} {[}ATLAS Collaboration{]},
 Phys.\ Lett.\ B \textbf{716}, 1 (2012) {[}arXiv:1207.7214 {[}hep-ex{]}{]}.


\bibitem{Higgs_CMS} S.~Chatrchyan \textit{et al.} {[}CMS Collaboration{]},
 Phys.\ Lett.\ B \textbf{716}, 30 (2012) {[}arXiv:1207.7235 {[}hep-ex{]}{]}.


\bibitem{comprehensive}
  S.~Chang, S.~K.~Kang, J.~-P.~Lee, K.~Y.~Lee, S.~C.~Park and J.~Song,
  arXiv:1210.3439 [hep-ph].


\bibitem{higgs hunter's guide} J.~F.~Gunion, H.~E.~Haber, G.~L.~Kane
and S.~Dawson, 
 Front.\ Phys.\ \textbf{80}, 1 (2000). 





\bibitem{2hdm_review} G.~C.~Branco, P.~M.~Ferreira, L.~Lavoura,
M.~N.~Rebelo, M.~Sher and J.~P.~Silva, 
 Phys.\ Rept.\ \textbf{516}, 1 (2012) {[}arXiv:1106.0034 {[}hep-ph{]}{]}.





\bibitem{LEP Higgs} {[}LEP Higgs Working Group for Higgs boson searches
and ALEPH and DELPHI and L3 and OPAL Collaborations{]}, 
 hep-ex/0107031. 

\bibitem{Tevatron}
A.~Abulencia {\it et al.} [CDF Collaboration], 
  \PRL96, 042003 (2006) [hep-ex/0510065]; V.~M.~Abazov {\it et al.} [D0 Collaboration], \PLB 682, 278 (2009)
  [arXiv:0908.1811 [hep-ex]]; T.~Aaltonen {\it et al.} [CDF Collaboration], 
  \PRL103, 101803 (2009) [arXiv:0907.1269 [hep-ex]]; V.~M.~Abazov {\it et al.} [D0 Collaboration], \PRD 80, 051107 (2009) 
  [arXiv:0906.5326 [hep-ex]].
  
\bibitem{typeII_constraints}
  O.~Deschamps, S.~Descotes-Genon, S.~Monteil, V.~Niess, S.~T'Jampens and V.~Tisserand,
  Phys.\ Rev.\ D {\bf 82}, 073012 (2010)
  [arXiv:0907.5135 [hep-ph]].


\bibitem{btosgamma1} V.~D.~Barger, M.~S.~Berger and R.~J.~N.~Phillips,
 Phys.\ Rev.\ Lett.\ \textbf{70}, 1368 (1993) {[}hep-ph/9211260{]}. 
S.~-S.~Bao, F.~Su, Y.~-L.~Wu and C.~Zhuang,
  Phys.\ Rev.\ D {\bf 77}, 095004 (2008)
  [arXiv:0801.2596 [hep-ph]].


\bibitem{PDG2012} J.~Beringer \textit{et al.} {[}Particle Data Group
Collaboration{]}, 
 Phys.\ Rev.\ D \textbf{86}, 010001 (2012). 


\bibitem{collider higgs} E.~Eichten, I.~Hinchliffe, K.~D.~Lane
and C.~Quigg, 
 Rev.\ Mod.\ Phys.\ \textbf{56}, 579 (1984) {[}Addendum-ibid.\ \textbf{58},
1065 (1986){]}. 
 N.~G.~Deshpande, X.~Tata and D.~A.~Dicus, 
 Phys.\ Rev.\ D \textbf{29}, 1527 (1984). 
 S.~S.~D.~Willenbrock, 
 Phys.\ Rev.\ D \textbf{35}, 173 (1987). 
 A.~Krause, T.~Plehn, M.~Spira and P.~M.~Zerwas, 
 Nucl.\ Phys.\ B \textbf{519}, 85 (1998) {[}hep-ph/9707430{]}. 


\bibitem{collider higgs2} A.~A.~Barrientos Bendezu and B.~A.~Kniehl,
 Phys.\ Rev.\ D \textbf{59}, 015009 (1999) {[}hep-ph/9807480{]}; S.~Moretti and K.~Odagiri, 
 Phys.\ Rev.\ D \textbf{59}, 055008 (1999) {[}hep-ph/9809244{]}; S.~-S.~Bao, X.~Gong, H.~-L.~Li, S.~-Y.~Li and Z.~-G.~Si,
  Phys.\ Rev.\ D {\bf 85} (2012) 075005
  [arXiv:1112.0086 [hep-ph]].



\bibitem{dominate process} J.~F.~Gunion, H.~E.~Haber, F.~E.~Paige,
W.~-K.~Tung and S.~S.~D.~Willenbrock, 
 Nucl.\ Phys.\ B \textbf{294}, 621 (1987). 
 R.~M.~Barnett, H.~E.~Haber and D.~E.~Soper, 
 Nucl.\ Phys.\ B \textbf{306}, 697 (1988). 
 F.~I.~Olness and W.~-K.~Tung, 
 Nucl.\ Phys.\ B \textbf{308}, 813 (1988). 


\bibitem{tripleb} V.~D.~Barger, R.~J.~N.~Phillips
and D.~P.~Roy, 
 Phys.\ Lett.\ B \textbf{324}, 236 (1994) {[}hep-ph/9311372{]};
  J.~F.~Gunion,
  Phys.\ Lett.\ B {\bf 322}, 125 (1994)
  [hep-ph/9312201];
  A.~Czarnecki and J.~L.~Pinfold,
  Phys.\ Lett.\ B {\bf 328}, 427 (1994)
  [hep-ph/9402212].
\bibitem{fourb}
  D.~J.~Miller,  S.~Moretti, D.~P.~Roy and W.~J.~Stirling,
  Phys.\ Rev.\ D {\bf 61}, 055011 (2000)
  [hep-ph/9906230];
  
  \bibitem{pt}
S.~Moretti, D.P.~Roy,
\PLB 470, 209 (1999)
[hep-ph/9909435];

\bibitem{dominate process3} S.~Moretti, K.~Odagiri, 
\PRD 55, 5627 (1997) [hep-ph/9611374];
C.~S.~Huang and S.~-H.~Zhu, 
 Phys.\ Rev.\ D \textbf{60}, 075012 (1999) {[}hep-ph/9812201{]}.


\bibitem{Kidonakis:2004ib} N.~Kidonakis, 
 JHEP \textbf{0505}, 011 (2005) {[}hep-ph/0412422{]}. 




\bibitem{H+Jetsub}
  S.~Yang and Q.~-S.~Yan,
  JHEP {\bf 1202}, 074 (2012)  [arXiv:1111.4530 [hep-ph]].  

\bibitem{NLO} 
  T.~Plehn,
  Phys.\ Rev.\ D {\bf 67}, 014018 (2003)
  [hep-ph/0206121];
  S.~-h.~Zhu,
  Phys.\ Rev.\ D {\bf 67}, 075006 (2003)
  [hep-ph/0112109];
  C.~Weydert, S.~Frixione, M.~Herquet, M.~Klasen, E.~Laenen, T.~Plehn, G.~Stavenga and C.~D.~White,
  Eur.\ Phys.\ J.\ C {\bf 67}, 617 (2010)
  [arXiv:0912.3430 [hep-ph]];M.~Klasen, K.~Kovarik, P.~Nason and C.~Weydert,
  Eur.\ Phys.\ J.\ C {\bf 72}, 2088 (2012)
  [arXiv:1203.1341 [hep-ph]].


\bibitem{Gopalakrishna:2010xm} S.~Gopalakrishna, T.~Han, I.~Lewis,
Z.~G.~Si and Y.~-F.~Zhou, 
 Phys.\ Rev.\ D \textbf{82}, 115020 (2010) {[}arXiv:1008.3508 {[}hep-ph{]}{]};
  S.~D.~Rindani and P.~Sharma,
  JHEP {\bf 1111}, 082 (2011)
  [arXiv:1107.2597 [hep-ph]].

\bibitem{Baglio:2011ap} 
  J.~Baglio, M.~Beccaria, A.~Djouadi, G.~Macorini, E.~Mirabella, N.~Orlando, F.~M.~Renard and C.~Verzegnassi,
  Phys.\ Lett.\ B {\bf 705}, 212 (2011)
  [arXiv:1109.2420 [hep-ph]];
  K.~Huitu, S.~Kumar Rai, K.~Rao, S.~D.~Rindani and P.~Sharma,
  JHEP {\bf 1104}, 026 (2011)
  [arXiv:1012.0527 [hep-ph]];
  




\bibitem{2HDM_constraints}
  F.~Mahmoudi and O.~Stal,
  Phys.\ Rev.\ D {\bf 81}, 035016 (2010)
  [arXiv:0907.1791 [hep-ph]].

\bibitem{Hewett:1992is}
  J.~L.~Hewett,
  Phys.\ Rev.\ Lett.\  {\bf 70}, 1045 (1993)
  [arXiv:hep-ph/9211256].
\bibitem{Barger:1992dy}
  V.~D.~Barger, M.~S.~Berger and R.~J.~N.~Phillips,
  Phys.\ Rev.\ Lett.\  {\bf 70}, 1368 (1993)
  [arXiv:hep-ph/9211260].
\bibitem{Bertolini:1990if}
  S.~Bertolini, F.~Borzumati, A.~Masiero and G.~Ridolfi,
  Nucl.\ Phys.\  B {\bf 353}, 591 (1991).


\bibitem{Brandenburg:2002xr} A.~Brandenburg, Z.~G.~Si and P.~Uwer,
 Phys.\ Lett.\ B \textbf{539}, 235 (2002) {[}hep-ph/0205023{]}.
 A.~Czarnecki, M.~Jezabek and J.~H.~Kuhn, 
 Nucl.\ Phys.\ B \textbf{351}, 70 (1991). 


\bibitem{Pumplin:2002vw} J.~Pumplin, D.~R.~Stump, J.~Huston,
H.~L.~Lai, P.~M.~Nadolsky and W.~K.~Tung, 
 JHEP \textbf{0207}, 012 (2002) {[}arXiv:hep-ph/0201195{]}. 


\bibitem{atlas0901}
  G.~Aad {\it et al.}  [ATLAS Collaboration],
  arXiv:0901.0512 [hep-ex].


\bibitem{Madevent} J.~Alwall, M.~Herquet, F.~Maltoni, O.~Mattelaer
and T.~Stelzer, 
 JHEP \textbf{1106}, 128 (2011) {[}arXiv:1106.0522 {[}hep-ph{]}{]}.


\bibitem{bbsu}
W.~Bernreuther, A.~Brandenburg, Z.~G.~Si and P.~Uwer,
  Phys.\ Lett.\ B {\bf 509},53 (2001)
  [hep-ph/0104096].
W.~Bernreuther, A.~Brandenburg, Z.~G.~Si and P.~Uwer,
  Nucl.\ Phys.\ B {\bf 690},81 (2004)
  [hep-ph/0403035].
W.~Bernreuther, A.~Brandenburg, Z.~G.~Si and P.~Uwer,
  Phys.\ Rev.\ Lett.\  {\bf 87},242002 (2001)
  [hep-ph/0107086].


\bibitem{breuthersi}
  W.~Bernreuther and Z.~-G.~Si,
  Nucl.\ Phys.\ B {\bf 837},90 (2010)
  [arXiv:1003.3926 [hep-ph]].


\end{thebibliography}
\end{document}